# Response to "Worrying Trends in Econophysics"

Joseph L. McCauley
Physics Department
University of Houston
Houston, Tx. 77204
jmccauley@uh.edu

and

Senior Fellow
**COBERA**
Department of Economics
*J.E.Cairnes Graduate School of Business and Public Policy*
NUI Galway, Ireland

Abstract

This article is a response to the recent "Worrying Trends in Econophysics" critique written by four respected theoretical economists [1]. Two of the four have written books and papers that provide very useful critical analyses of the shortcomings of the standard textbook economic model, neo-classical economic theory [2,3] and have even endorsed my book [4]. Largely, their new paper reflects criticism that I have long made [4,5,6,7,] and that our group as a whole has more recently made [8]. But I differ with the authors on some of their criticism, and partly with their proposed remedy.

# 1. Money, conservation laws, and neo-classical economics

*Our concerns are fourfold. First, a lack of awareness of work which has been done within economics itself. Second, resistance to more rigorous and robust statistical methodology. Third, the belief that universal empirical regularities can be found in many areas of economic activity. Fourth, the theoretical models which are being used to explain empirical phenomena.*

*The latter point is of particular concern. Essentially, the models are based upon models of statistical physics in which energy is conserved in exchange processes.*

<div style="text-align: right">Gallegati, Keen, Lux, and Ormerod [1]</div>

Since the authors of "Worrying" [1] begin with an attack on assumptions of conservation of money and analogs of conservation of energy in economic modelling, let me state from the outset that it would generally be quite useless to assume conserved quantities in economics and finance (see [4], Ch. 1). There is no reliable analog of energy in economics and there are very good reasons why no meaningful thermodynamic analogy can be constructed (see [4, Ch. 7]). In particular, money is not conserved. Money is created and destroyed rapidly via credit. Leveraging leads to big changes in 'money'. Are there any conservation laws at all in real markets, and if so do they have any significance for deducing market dynamics?

Conservation laws follow from invariance principles [8], so one should not expect conservation laws in socio-economic 'motions' like financial transactions or production and consumption [4,5]. We have identified exactly one invariance principle in finance: no arbitrage is equivalent to a discrete version of rotational invariance of the price distribution [see [4], Ch. 7). For inviolable mathematical laws of motion rather then era-dependent mathematical modelling we know from Wigner's explanation why mathematics has been so 'unreasonably effective' in physics [8] that we would need the equivalent of all four space-time invariance principles: translational invariance, rotational invariance, time translational invariance, and Galilean invariance. Those local invariance principles are the foundation for the discovery of mathematical law in classical mechanics, quantum mechanics, and general relativity, as Wigner has explained. Wigner pointed out that without those underlying invariance principles mathematical law might well exist, in principle, but the world would be too complicated for us to discover mathematical law in practice. That is exactly where we stand in trying to make a science of socio-economic phenomena [7], but even in the absence of inviolable mathematical laws of trading behavior we are still able to generate finance market statistics dynamically accurately [4] in our era.

In the definition of the neo-classical economic model ('general equilibrium theory' [9]) one global conservation law is assumed at the start. However, for the global integrability presumed for the existence of equilibrium there actually must be n global conservation laws, where n is the number of different goods exchanged by agents. The budget constraint combined with price dynamics, $dp/dt=D(p)-S(p)$ (where D is total demand at price p and S is the total supply), yields Walras' law, a global conservation law that puts the motion in price space (phase space) on an n-sphere [4,9]. The budget constraint *is* conservation of money, and that constraint is badly violated in the real world, where money is created and destroyed with the tap of a computer key via credit.

In the beginning of our current era of credit and deregulation, which we can date from the early 1970s, the supply-demand curves predicted by neo-classical economics [9] were falsified by Osborne [10], who simultaneously predicted the observable microeconomic supply and demand curves. In a strict gold-standard economy money would be conserved if mining production were ignored. Credit is required for the expansion of both industry and consumption, which is why we're no longer on any sort of gold standard. Nixon took the Dollar off the (Rooseveltian) gold standard in 1972 due to the inflation created by the Vietnam War. Option pricing was deregulated in 1973, accidentally parallel with the revolutionary Black-Scholes paper. OPEC correctly understood the Dollar devaluation and raised the price of oil dramatically to compensate for the rapid inflation that would follow the degradation of the dollar. The deregulation of financial markets that followed is described in Liar's poker [11].

In contrast with the main concern of "Worries" [1], the work done in Fribourg, Boston, Alessandria, UCLA, Palermo, Oldenburg, and Houston, to name a few centers of econophysics research, do not assume conservation of money or any analog of conservation of energy. Searches for conservation laws can be found in the neo-classical economics literature [12,13], but those efforts bore no fruit at all for lack of a firm basis in empirical data.

There is one econophysics model assuming conservation of money that is interesting within the historical context of neo-classical economic theory. Radner had observed in 1968 [14] that money/liquidity cannot appear in neo-classical economics: with perfect foresight, perfect planning into the infinite future, there can be no liquidity demand (in real markets we have, in contrast, largely total ignorance, market as pure nonGaussian noise). Radner speculated that money/liquidity arises from uncertainty and also computational limitations, although he did not say what he meant by "computation" and apparently had no idea at all of a Turing machine [15]. Conservation of money is assumed in the simple trading model made by Bak, Nørrelyke and Shubik [16]. There, expected utility was maximized (so the paper is neo-classsical) and then noise was added to see if "money" would emerge from the dynamics of trading. The attempt failed, the model is empirically irrelevant (as is every model of optimizing behavior) but can be understood as an honorable effort to address the mathematical problem posed by Radner's discussion. I have no evidence that Per Bak was aware of Radner's paper, but maybe Shubik was aware of it. The Bak et al model is asymptotically a pure barter model because statistical equilibrium is reached, and in general the authors of "Worries" are right that making further barter models is generally (both empirically and theoretically seen) useless. In our era, conservation of money is a silly assumption. The paper by Bak et al tried to bridge the gap between standard economic theory and econophysics but was too ambitious, the gap is a chasm.

## 2. Should econophysicists rely on the economics literature for guidance?

*Our concerns about developments within econophysics arise in four ways:*

- *a lack of awareness of work which has been done within economics itself*
- *resistance to more rigorous and robust statistical methodology*
- *the belief that universal empirical regularities can be found in many areas of economic activity*
- *the theoretical models which are being used to explain empirical phenomena*

*…. One of the present authors has drawn the analogy of imagining 'a new paper on quantum physics with no reference to received literature' […].*

<div style="text-align:right">Gallegati, Keen, Lux, and Ormerod [1]</div>

This leads into another concern expressed in [1], that ideas of production in economics theory are ignored by econophysicists. The problem with that criticism is that neo-classical models of production are no better than neo-classical exchange models: there is inadequate or nonexisting empirical basis for any neo-classical assumption. Econophysicists are safer to ignore the lessons taught in standard economics texts (both micro- and macro-) than to learn the economists' production ideas and take them seriously (see Dosi [17] for non optimizing models of production, and see Zambelli [18] for a thorough and very interesting study of the degree of computational complexity of neo-classical production models). To be quite blunt, all existing 'lessons' taught in standard economics texts should be either abandoned or tested empirically, but should never be accepted as a basis for modelling. It's important to understand what economists have done previously mainly to avoid making the same mistakes. Toward that end, see also Velupillai [19] for a critical discussion of the neo-classical equilibrium model from an economists' perspective, where lack of computability of the neo-classical model would prevent agents from locating equilibrium even if the equilibrium point were stable, and even if the agents were as unrealistically well informed as the standard model so uncritically presumes.

I will explain below why statistical analyses performed from the economists' standpoint are generally untrustworthy, with few exceptions [20]. Our alternative to taking advice from economics texts and papers for the purpose of modelling markets will be stated below.

Osborne [21] first introduced the lognormal stock pricing model in 1958 and should be honored as the first econophysicist, but even econophysicsts cite Bachelier while ignoring Osborne. Bachelier independently of Einstein, Markov and Smoluchowski produced some of the first results on Markov processes, but his finance model predicted negative prices and did not describe any market correctly, even to zeroth order. Econophysicists have contributed mightily in recent years to the study of financial markets, but that work is not yet finished. The authors of [1] are correct that the so-called 'rigorous and robust' methods of statisticians have generally not been used in econophysics, and for good reasons. In our group we begin with market histograms and then deduce an empirical model [4,21].

We have no mathematical model in mind a priori. We do not 'massage' the data. Data massaging is both dangerous and misleading. Econometricians mislead themselves and others into thinking that their models help us to understand market behavior. The reason is that economists assume a preconceived model [22,23] with several unknown parameters, and then try to force fit the model to a nonstationary time series by a 'best choice of parameters' (using 'rigorous and robust methods'). Their conclusion is that the data are too hard to fit over long time scales [23]. Because we deduce our model from the data [4,7,24], we arrive at exactly the opposite conclusion: due to nonuniqueness in fitting any infinite precision model (stochastic or deterministic model) to inherently finite precision data, it is too easy to fit any set of histograms constructed from time series. We've illustrated this fact for financial time series but the lesson of nonuniqueness is not new: we anticipated it from our experience in nonlinear dynamics [4,5]. The reason why our group has not yet ventured into nonfinancial economic data analysis is that the problem there is horrendous: because of the sparseness of nonfinancial market data, it will likely be possible to fit the data using too many different unrelated classes of models [4,7]. The best one can hope for there will be to rule out certain model classes. One will never be able to 'understand' the market data the way that we've managed to 'understand' financial markets.

There is another very good reason to concentrate on finance markets in this era. With the magnitude and frequency of the enormous daily and even second by second transfers, finance markets dominate all other markets. Finance markets drive other markets globally. For some reason that I fail to understand, economists still tend to see finance markets as only one relatively insignificant market in their panoply.

*It is easy to generate hypothetical data sets which, when binned, can give misleading pictures of the statistical distributions which characterise them. [...] shows that data generated from a lognormal distribution can be easily misinterpreted as evidence of power laws.*
Gallegati, Keen, Lux, and Ormerod [1]

It's an old fractal-wive's tale from the 1980s that lognormal densities mimic fat tails if you merely 'eyeball' them [25]. In a careful analysis using logarithmic returns, no such crude error is made. The reader must be aware that, with few enough decades in a log-log plot, $\ln f(x)$ looks linear in $\ln x$ where $f(x)$ is a large class of smooth functions. About nonuniqueness in fitting infinite precision models to finite precision data, we have written enough [4,5,7].

However, the authors of [1] are correct that statistical physicists often express an unjustified expectation for universal scaling exponents even when there is no evidence for a critical point/bifurcation [4,5]. There is apparently no universality of scaling exponents in finance, where the fat tail exponents range from roughly 2 to 7 [26] and are market dependent. Moreover, there is no reason at all to expect universality in socio-economic behavior [4,5].

There is other criticism that can be levelled at econophysics and economics alike: most of us have not read and understood Osborne's 1977 falsification [10] of standard economics theory, and so are not aware that we can expect only misguidance from most economics texts and papers. In econophysics, agent based modelling is often performed in an effort to explain fat tails, but fat tails without other market constraints are too easy to 'explain' [4]. Again, the authors of "Worries" are right that background in statistical physics has generally led to unjustified emphasis on scaling and universality. Agent based models have been constructed where the motion is asymptotically stationary, but stationary markets do not exist (I give no references here in order to protect the guilty, some of whom are my friends).

Markets are not merely nonstationary, the time series generally exhibit nonstationary increments [27] as well, meaning that x(t+T)-x(t)≠x(T), where by equality we mean 'equality in distribution' (in addition, x(t) is assumed defined so that x(0)=0). This means that the increments depend not merely on the time difference T but on the starting point t as well. *This complication makes time series analysis devilishly tricky*. We are not yet satisfied that we understand how to perform a noncircular or mistake-free analysis of finance data that searches for scaling behavior [28]. The scaling of interest to us is of the form $f(x,t)=t^{-H}F(x/t^H)$, where x is the logarithmic return and f is the returns density. That is, we look for a data collapse of the form $F(u)=t^H f(x,t)$ where $u=x/t^H$. If such scaling holds, then the variance scales as $t^H$, but that alone is not evidence for market correlations [27]. Papers have been published by very respectable finance researchers claiming market correlations on the basis of a Hurst exponent H≠1/2, see e.g. [29], but a Hurst exponent H≠1/2 does not imply long time correlations unless it has been first established that the time series has stationary increments [27] (see also [30,31] for a discussion of the scaling dynamics and lack of autocorrelations in Levy models). These different ideas are usually confused together into a thick soup by physicists and economists alike. In particular, one should question any paper claiming to deduce a Hurst exponent from a spectral density: nonstationary time series have no spectral decomposition, and all financial time series are nonstationary (the moments of prices p(t) or returns $x(t)=\ln p(t)/p_c$ never approach constants, where $p_c$ is dynamically defined 'value' [24]). Further, a nonstationary time series with unknown dynamics cannot be transformed into a stationary one (try it, as an exercise, using Ito calculus). The transformations purported to perform such magic in econometrics are akin to wishful thinking or praying for rain. Our dynamic definition of value is new and applies to nonfinancial markets as well, but has so far been ignored by economists.

## 3. Was econophysics previously anticipated?

*Econophysics has already made a number of important empirical contributions to our understanding of the social and economic world. Many of these were anticipated in two truly remarkable papers written in 1955 by Simon ... and in 1963 by Mandelbrot ... , the latter in a leading economic journal.*

<div style="text-align: right;">Gallegati, Keen, Lux, and Ormerod [1]</div>

Ignoring Osborne altogether, Simon and Mandelbrot are named [1] as having anticipated the main results of econophysics. This is false. Mandelbrot's main contributions were the introduction of Levy distributions (fat tails for tail exponents in the range from 1 to 3) [32], the discussion of efficient markets in finance [33], and his explicit model of fractional Brownian motion (fBm) [34].

Real financial markets are very hard to beat. The efficient market hypothesis (EMH) states that markets are impossible to beat: there are no correlations, no systematic patterns in prices that can be exploited for profit. If the drift can be subtracted out of a returns time series, then financial markets are Markovian (pure nonGausssian noise) to zeroth order, meaning that correlations like fBm would have to be a higher order effect [27]. Mandelbrot did not anticipate the correct description of the dynamics of financial markets, which is diffusive [4,27] rather than Levy [30,31]. Our modelling is a generalization of Osborne's Gaussian returns model to the case where the diffusion coefficient $D(x,t)$ is $(x,t)$ dependent. Osborne's model satisfies the EMH as does every Markov model.

Fat tails in real markets are generally not described by Levy distributions because the Levy exponent range is from 1 to 3, whereas market exponents range from 2 to 7. We can generate the observed fat tail densities via our approach, including all exponents from 2 to infinity [27] and with finite variance for exponents from 3 to infinity. But fat tails are not the main feature of interest in financial market dynamics. E.g., option prices blow up if fat tails are included [24,35]. Option traders must anticipate the dynamics of markets for small to moderate returns, but VAR requires the entire range of the returns distribution.

So far as I can see, Herbert Simon anticipated no quantitative results from econophysics. Simon wrote many words but produced no mathematical market model. His paper on asymmetric distributions [1] is not an anticipation of econophysics. Levy distributions are generally asymmetric, exponential distributions are generally asymmetric, so are nearly all distributions. The trick is to generate the empirically observable market distributions dynamically, and Simon did not do that. His main idea, 'Bounded Rationality', can be interpreted as constraint on optimizing behavior (albeit not necessarily leading to equilibrium) and is therefore still a neo-classical rather than an empirically based idea, although behavioral economists have tried through experiments to ground 'bounded rationality' empirically. The Bak et al model [16] can be seen as one of the few mathematically specific bounded rationality models (see also [36,37]). So far as I know, bounded rationality has produced no falsifiable predictions whereas econophysics, like physics, is based on on falsifiable predictions. But the mainstream in the economics profession did not follow Simon, they instead followed game theory via Nash, whose 'every man for himself' equilibria they found to be comfortably neo-classical in spirit [38]. So far, that approach has not managed the transition from the nightmares of Dr. Strangelove [39] to the empirical reality of generating realistic market statistics or explaining markets qualitatively. But Nash's game is not the only game in town.

It's well known that Brian Arthur inspired the Minority Game, which indeed has been solved extensively using analysis from statistical physics [40]. The game theoretical approach does not describe money transactions in real finance markets quantitatively, but some members of the Fribourg school have also made interesting agent based trading models based on actual limit book data [41]. We know now that an agent based model should reproduce at least three aspects of real markets in order to be taken seriously: nonstationarity with nonstationary increments, volatility (described locally by an (x,t) dependent diffusion coefficient, characterizing the noise traders), and fat tails. Predictions of fat tails from stationary distributions are both easy and uninteresting.

In our group we've found that statistical physics alone does not provide an adequate background for econophysics. The advanced methods of statistical physics are confined to motions near statistical equilibrium whereas real financial markets exhibit no equilibrium whatsoever. *Empirical evidence for statistical equilibrium has never been produced for any real market*. We've found our experience in nonlinear dynamics to be very useful for thinking about the dynamics of real markets, although finance markets are not at all approximately deterministic. For example, the distinction between local and global behavior is stressed in nonlinear dynamics, and one understands better why 'general equilibrium theory' (neo-classical economics) is an impossible candidate for describing real markets if one has background in nonlinear dynamics (where integrability vs. nonintegrability and local vs. global are the key ideas). The stochastic models that we've discovered [27,42] in our (unending) attempt to understand financial markets cannot be found in any physics or math text. They were anticipated by no one, including Mandelbrot, Osborne, and Simon. They constitute an entirely new contribution to the theory of Markov processes. Finally, we place little or no weight on the related notions of 'data mining' and 'financial engineering'. Nothing fundamental about how markets work has been discovered by following either path.

Neo-classical economics has been falsified [10]. That it's still the central topic of economics texts means that it can be understood as the mathematized ideology of our era [4]. In physics and econophysics there are also trends that can be understood largely as mathematized philosophy. One of these is the Tsallis movement. Initially, the Tsallis model was proposed ad hoc as a statistical equilibrium model. Then, it was generalized to a dynamical model, but the original equilibrium model cannot be reproduced by the dynamic one applied in finance [43,44] with Hurst exponent H=1/(3-q), where q is the index in the model. Worse, the Tsallis dynamics model was presented for years as 'nonlinear' and nonMarkovian [43,44], while at the same time implicitly assuming a Markovian description in the form of an equivalent Langevin equation. No one in the Tsallis movement bothered to ask if a nonlinear diffusion equation is consistent with a Langevin equation. A Langevin equation with ordinary functions (rather than functionals) as drift and diffusion coefficients is always Markovian [27]. A Langevin description of truly nonlinear diffusion is impossible because a nonlinear diffusion equation has no Green function, and the Green function (transition probability density) is the defining feature of a Markov process. We showed recently that the Tsallis model is merely one special case of Markov dynamics, linear diffusion, for the general class of quadratic diffusion coefficients that scale with a Hurst exponent [27]. For the entire class of scaling quadratic diffusion coefficients a 'nonlinear disguise' is possible for the probability density, but the motion is generated by a (necessarily linear!) Fokker-Planck equation and so is strictly Markovian. Contrary to all the unjustified claims in the literature [43,44], there is no such thing as a

'nonlinear Fokker-Planck equation'. Mathematized philosophy like neo-classsical economic theory and the Tsallis model drain energy and effort away from meaningful research.

*... the 'log-periodic' business had been exaggerated into doomsday speculations published in serious journals [... for example], although not even its prophecies for the mundane world of the stock markets had ever been scrutinized in a rigorous fashion.*

The logperiodic model [45] is another model from econophysics. It gets both a plus and some minuses from us. The plus side is that the idea of a finite time singularity is interesting and should either be established or else falsified by a more careful and more critical empirical analysis. This leads to the minus side: there is no reason in that model (or in real data) to claim a bifurcation or critical point, all that appears in the model is a finite time singularity. Unfortunately, the model as constructed is not falsifiable: it has far too many free parameters.

## 4. Where are we headed?

> *'I believe in God and I believe in free markets."*
> Enron CEO Kenneth Lay,
> San Diego Union-Tribune, 2001

There is little or nothing in existing micro- or macroeconomics texts [9,46,47] that is of value for understanding real markets. Economists have not understood how to model markets mathematically in an empirically correct way. Their only (once-) scientific model so far, the neo-classical one, has been falsified. What is now taught as standard economic theory will eventually disappear, no trace of it will remain in the universities or boardrooms because it simply doesn't work [48,49]: were it engineering, the bridge would collapse. The reigning economic theory, based on the assumptions of optimizing behavior with infinite foresight, will be displaced by econophysics (where noise rather than foresight reigns supreme), meaning empirically based modelling where one asks not what we can do for the data (give it a massage), but instead asks what can we learn from the data about how markets really work. Existing standard economics texts are filled with scads of graphs [46,47], but those graphs are merely cartoons because they don't represent real data, they represent only the falsified expectations of neo-classical equilibrium theory [4]. No existing economic model provides us with a zeroth order starting point for understanding how real markets function. No current economics model or idea provides a starting point for building an interesting falsifiable market model. The future economics theory/econophysics texts will be filled with histograms, time series, and with graphs with big error bars, all representing real markets. They will look more like nuclear physics data (few data points, big error bars) than like the smooth curves in the existing economics texts. The aim will be to try to understand how markets work, not to present an irrelevant model about how a real market ideally should (but cannot) behave. Comparisons of neo-classical economics with Newton's first law are irrelevant and terribly misleading: Newton's first law is realized locally in every observable gravitational trajectory. Neo-classical behavior, in contrast, approximates no real market locally. Future economics texts will ask many questions but will provide few answers. That is uncomfortable, but that is the reality of markets. The good news is that economics research will not end because most of the important questions there can likely never be answered. Soros' message [49] must be taken seriously and understood because he's right. In social behavior, each agent has a perception/illusion of reality but one never knows how big is the gap between one's illusion and 'social reality', which is always unknown. We get an incomplete picture of social reality from history (market time series). As time goes on, one gains new knowledge and has the chance to correct one's perception of reality, one has the chance partly to correct one's mistakes [49], but our illusions can never completely or in controlled fashion be corrected. Skepticism, not belief, must reign supreme in economics, finance, and politics. Soros has repeated his basic ideas in too many books, but we can see him as one of the most perceptive social scientists of our era. No one knows yet how to mathematize what Soros tries to teach us about markets [24]. Our empiric-dynamic identification of 'value' is the starting point for thinking about Soros' type of 'technical analysis'. To try to follow Soros' argument and mathematize it is to deal with true mathematical complexity, something that we have done neither in economics nor in econophysics (see Zambelli [18] for an example of how to program a Turing machine and study the algorithmic complexity of a model on your laptop). True complexity has been handled so far only in biology (the genetic code and its consequences), where, as Ivar Giæver has noted [50], there are many facts and very few equations.

Let me end anecdotally and then with a suggestion. Physicists have gone into biology and econophysics because that's where the interesting problems are.

Turbulence and quantum coherence (mesoscopics) remain the outstanding unsolved problems of physics. We meet real complexity in nature in cell biology, also in quantum computation, along with econophysics they're the fields of the future in physics. Every physics student should be required to take a good, stiff course in cell biology from a text like „Fat Alberts" [51], as Ivar Giæver calls it. Physicists were hired by banks and trading houses beginning in the 1980s. It's no mistake that Fischer Black, the driving force behind the Black-Scholes-Merton model (which assumes Osborne's lognormal distribution without referencing it) was trained as a physicist. Physicists go into econophysics largely because it's interesting, and with the hope of getting a well-paying job after the Ph. D. Trading houses do not hire many people from economics or finance because the products of economics and business departments don't know enough math and are unprepared for both data analysis and computer intensive modeling. I gave a talk to the Enron modelers in June, 2000. Of the thirty, they were overwhelmingly from physics and engineering, with two from math. None were from economics. Two group leaders had Ph.D.s in experimental physics. The management, not the modellers, made Enron go belly up. George Soros, if he's true to his words, does not carry any belief that's free of skepticism. Enron wanted our group to work for them in 2000. We refused because they wanted data mining and also technical research aimed at pricing weather options. The offer of money was not seductive, we strongly and rightly preferred to stick with fundamental research of our own choice, so we refused. The Arrow-Debreu-Merton program [52] fails: when insufficient liquidity exists, meaning market statistics are too sparse or even nonexistent (as with gas stored in the ground, or predicting details of the weather). In such a case option pricing is like a stab in the dark, but Enron modelers were required by management to price options on gas stored in the ground. It was necessary to invent statistics out of thin air. You simply cannot put a price on everything when the noise traders [4, 53] are too few to provide liquidity.

The real problem with my proposal for the future of economics departments is that current economics and finance students typically do not know enough mathematics to understand (a) what econophysicists are doing, or (b) to evaluate the neo-classical model (known in the trade as 'The Citadel') critically enough to see, as Alan Kirman [54] put it, that 'No amount of attention to the walls will prevent The Citadel from being empty'. I therefore suggest that the economists revise their curriculum and require that the following topics be taught: calculus through the advanced level, ordinary differential equations (including advanced), partial differential equations (including Green functions), classical mechanics through modern nonlinear dynamics, statistical physics, stochastic processes (including solving Smoluchowski-Fokker-Planck equations), computer programming (C, Pascal, etc.) and, for complexity, cell biology. Time for such classes can be obtained in part by eliminating micro- and macro-economics classes from the curriculum. The students will then face a much harder curriculum, and those who survive will come out ahead. So might society as a whole.

**Acknowledgement**

I'm grateful to Gemunu Gunaratne, Kevin Bassler, and Cornelia Küffner for discussions and comments. I'm grateful to Rebecca Schmitt, Vela Velupillai, and Giulio Bottazzi for stimulating discussions of how to improve the current economics curriculum mathematically, and to a referee for information about different approaches to Herbert Simon's ideas.